%% file: a2657.tex
\documentclass[fleqn,usenatbib]{mnras}

\usepackage{newtxtext,newtxmath}

\usepackage[T1]{fontenc}

\DeclareRobustCommand{\VAN}[3]{#2}
\let\VANthebibliography\thebibliography
\def\thebibliography{\DeclareRobustCommand{\VAN}[3]{##3}\VANthebibliography}

\usepackage{graphicx}	
\usepackage{amsmath}	
\usepackage{orcidlink}
\usepackage[export]{adjustbox}

\input{latex_mycommands.txt}

\title[A radio bubble shredded by gas sloshing?]{A radio bubble shredded by gas sloshing?}

\author[A. Botteon et al.]{A. Botteon\orcidlink{0000-0002-9325-1567},$^{1}$\thanks{E-mail: \href{mailto:andrea.botteon@inaf.it}{andrea.botteon@inaf.it}},
F. Gastaldello\orcidlink{0000-0002-9112-0184},$^{2}$
J. A. ZuHone\orcidlink{0000-0003-3175-2347},$^{3}$
M. Balboni\orcidlink{0009-0001-3048-0020},$^{2,4}$
I. Bartalucci\orcidlink{0000-0001-7703-9040},$^{2}$
G. Brunetti\orcidlink{0000-0003-4195-8613},$^{1}$
\newauthor
A. Bonafede\orcidlink{0000-0002-5068-4581},$^{5,1}$
M. Br\"{u}ggen\orcidlink{0000-0002-3369-7735},$^{6}$
T. W. Shimwell\orcidlink{0000-0001-5648-9069}$^{7,8}$
and
R. J. van Weeren\orcidlink{0000-0002-0587-1660}$^{8}$
\\
$^1$INAF - IRA, via P.~Gobetti 101, I-40129 Bologna, Italy \label{ira} \\
$^2$INAF - IASF Milano, via A.~Corti 12, I-20133 Milano, Italy \label{iasf} \\
$^3$Center for Astrophysics, Harvard \& Smithsonian, 60 Garden St., Cambridge, MA 02138, USA \label{cfa} \\
$^4$DiSAT, Università degli Studi dell'Insubria, Via Valleggio 11, I-22100 Como, Italy \label{insburia} \\
$^5$Dipartimento di Fisica e Astronomia, Universit\`{a} di Bologna, via P.~Gobetti 93/2, I-40129 Bologna, Italy \label{unibo} \\
$^6$Hamburger Sternwarte, Universit\"{a}t Hamburg, Gojenbergsweg 112, D-21029 Hamburg, Germany \label{hamburg} \\
$^7$ASTRON, the Netherlands Institute for Radio Astronomy, Postbus 2, NL-7990 AA Dwingeloo, The Netherlands \label{astron} \\
$^8$Leiden Observatory, Leiden University, PO Box 9513, NL-2300 RA Leiden, The Netherlands \label{leiden} 
}

\date{Accepted XXX. Received YYY; in original form ZZZ}

\pubyear{2023}

\begin{document}
\label{firstpage}
\pagerange{\pageref{firstpage}--\pageref{lastpage}}
\maketitle

\begin{abstract}
We report on the detection of diffuse radio emission with peculiar morphology in the central region of the galaxy cluster Abell 2657. The most striking feature identified in our 144~MHz \lofar\ image is a bifurcated radio arc that extends for a projected size of 150$-$200~kpc. From the analysis of \xmm\ data, we find clear evidence of gas sloshing in the cluster and a possible dip in X-ray surface brightness between the two radio arcs which deserves confirmation. Interestingly, the synchrotron emission of the bifurcated radio arc is stretched along the sloshing spiral. We compare our observational results with numerical simulations of non-thermal components interacting with gas motions. We suggest that the detected emission may trace a radio bubble shredded by gas sloshing, where relativistic electrons and magnetic fields are expected to be stretched and stirred as a consequence of tangential flows induced by the spiralling gas motion. Lastly, we report on the presence of two thin (6$-$7~kpc in width) and parallel strands of radio emission embedded in the outer arc that are morphologically similar to the emerging population of non-thermal filaments observed in galaxy clusters, radio galaxies, and the Galactic centre. While this work further demonstrates the complex interplay between thermal and non-thermal components in the intracluster medium, follow-up observations in radio and X-rays are required to firmly determine the origin of the features observed in Abell 2657.
\end{abstract}




\section{Introduction}

Galaxy clusters, the largest gravitationally bound objects in the Universe, enable us to study the large-scale structure formation process and several aspects of plasma (astro)physics. Indeed, clusters are permeated by a hot and diffuse plasma, known as the intracluster medium (ICM), the thermal component of which can be observed in the X-ray band due to its thermal bremsstrahlung emission. This plasma contains signatures left by numerous astrophysical phenomena that occur during cluster formation and evolution. In particular, gas motions can be driven by major/minor mergers and by energetic outbursts of the active galactic nucleus (AGN) such as those associated with the brightest cluster galaxy (BCG), leading to the generation of shocks, cold fronts, and turbulence in the ICM \citep[\eg][for reviews]{markevitch07rev, simionescu19rev, zuhone22rev}. \\
\indent
The thermal gas in the ICM is known to have a strong interplay with the non-thermal components, namely relativistic electrons and magnetic fields, that can be studied in the radio band due to their synchrotron emission. For example, it is currently believed that turbulence and shocks in the ICM lead to the formation of diffuse cluster-scale radio sources, although the role of additional physical processes is still under investigation \citep[\eg][for reviews]{brunetti14rev, vanweeren19rev}. Another well-studied interaction between non-thermal components and the hot cluster atmosphere is probed by the displacement of thermal gas due to bubbles of radio plasma ejected by the central AGN, which leads to the establishment of radio lobes/X-ray cavities systems \citep[\eg][for reviews]{mcnamara07rev, mcnamara12rev, fabian12rev, gitti12rev}. \\
\indent
In recent years, the study of non-thermal phenomena in galaxy clusters and their relation with the thermal gas properties and dynamics has been enhanced thanks to highly-sensitive observations performed with modern radio interferometers. New data show a wealth of radio features with increasing complexity in the ICM that suggest a tight connection between outflows of cluster radio galaxies, thermal gas motions, and diffuse synchrotron sources. In particular, the non-thermal plasma ejected by AGN in clusters is believed to play a fundamental role in the formation of extended cluster radio emission as it may enrich the environment with seed relativistic electrons and magnetic fields that can be re-accelerated and amplified by shocks and turbulence in the ICM. The interaction between non-thermal components injected by radio galaxies and thermal gas motions in clusters has been the subject of recent magnetohydrodynamical (MHD) simulations \citep{vazza21transport, vazza23cycle, zuhone21transport, zuhone21bubbles, fabian22interaction}. These works indicate that non-thermal components (i) can be efficiently spread and fill a significant volume of the ICM during the cluster formation process and (ii) their emission can take a variety of morphologies in the radio band due to the complex interplay with thermal gas motions. In this paper, we report on the observation of a peculiar diffuse radio emission in the centre of a galaxy cluster that could represent one of such interplay. \\
\indent
Abell 2657 (RA: 23$^h$44$^m$51$^s$, Dec: +09\deg08\arcmin40\arcsec), hereafter A2657, is a galaxy cluster at $z=0.04$ \citep{smith04}. Due to its X-ray flux of $2.5\times10^{-11}$ \ergscmsq\ in the 0.1$-$2.4 keV band, it belongs to the HIghest X-ray FLUx Galaxy Cluster Sample \citep[HIFLUGCS;][]{reiprich02}. In \citet{hudson10}, it is classified as a weak cool-core cluster with a virial temperature of $3.5\pm0.1$ keV and an overall roundish morphology, although some weak indications of dynamical disturbance have been pointed out \citep[\eg][]{markevitch98asca}. A2657 appears in the second catalogue of \planck\ Sunyaev-Zel'dovich sources with the name PSZ2G096.77-50.29, where it is reported to have a mass within 500 times the critical density at the cluster redshift of $\mfive = (1.5 \pm 0.2) \times 10^{14}$ \msun\ \citep{planck16xxvii}, which makes it a relatively low-mass system to host cluster-scale diffuse radio emission. \\
\indent
In the following, we study the central region of A2657 by using recent \lofarE\ \citep[\lofar;][]{vanhaarlem13} observations and archival \xmm\ data. We complement our analysis by comparing the observational results with that from state-of-the-art numerical simulations. We assume a $\Lambda$ cold dark matter cosmology with $\omegal = 0.7$, $\omegam = 0.3$, and $\hzero = 70$ \kmsmpc\ throughout. At $z=0.04$, this corresponds to a luminosity distance of $D_{\rm L} = 176.5$ Mpc and to an angular scale of 0.791 kpc arcsec$^{-1}$. We adopt the convention $S_\nu \propto \nu^{-\alpha}$ for radio synchrotron spectrum, where $S_\nu$ is the flux density at frequency $\nu$ and $\alpha$ is the spectral index.

\begin{figure*}
 \centering
 \includegraphics[height=7.8cm,trim={0.1cm 0.2cm 0.1cm 0.2cm},clip,valign=c]{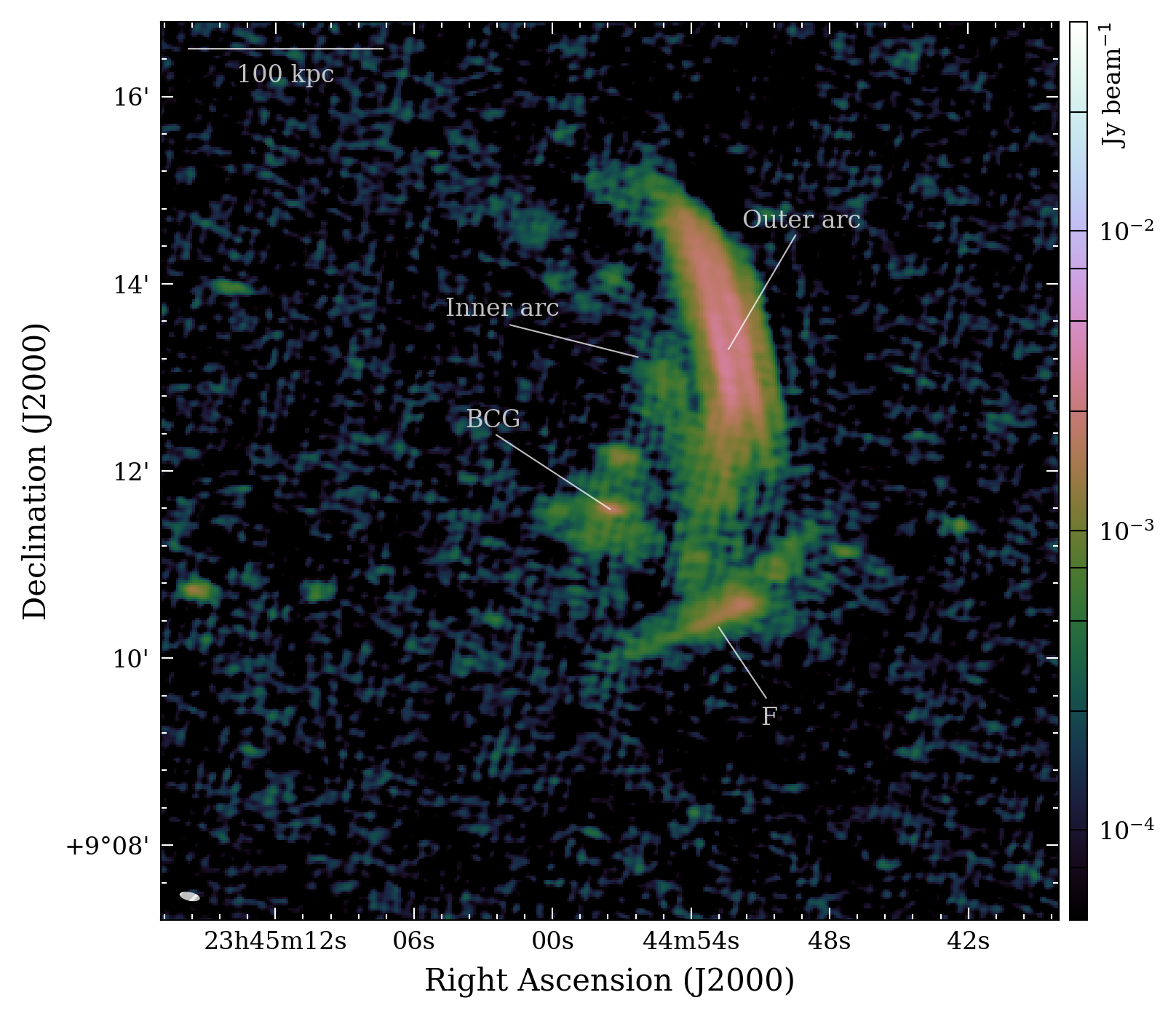} 
 \includegraphics[height=7.8cm,trim={0.1cm 0.2cm 0.1cm 0.2cm},clip,valign=c]{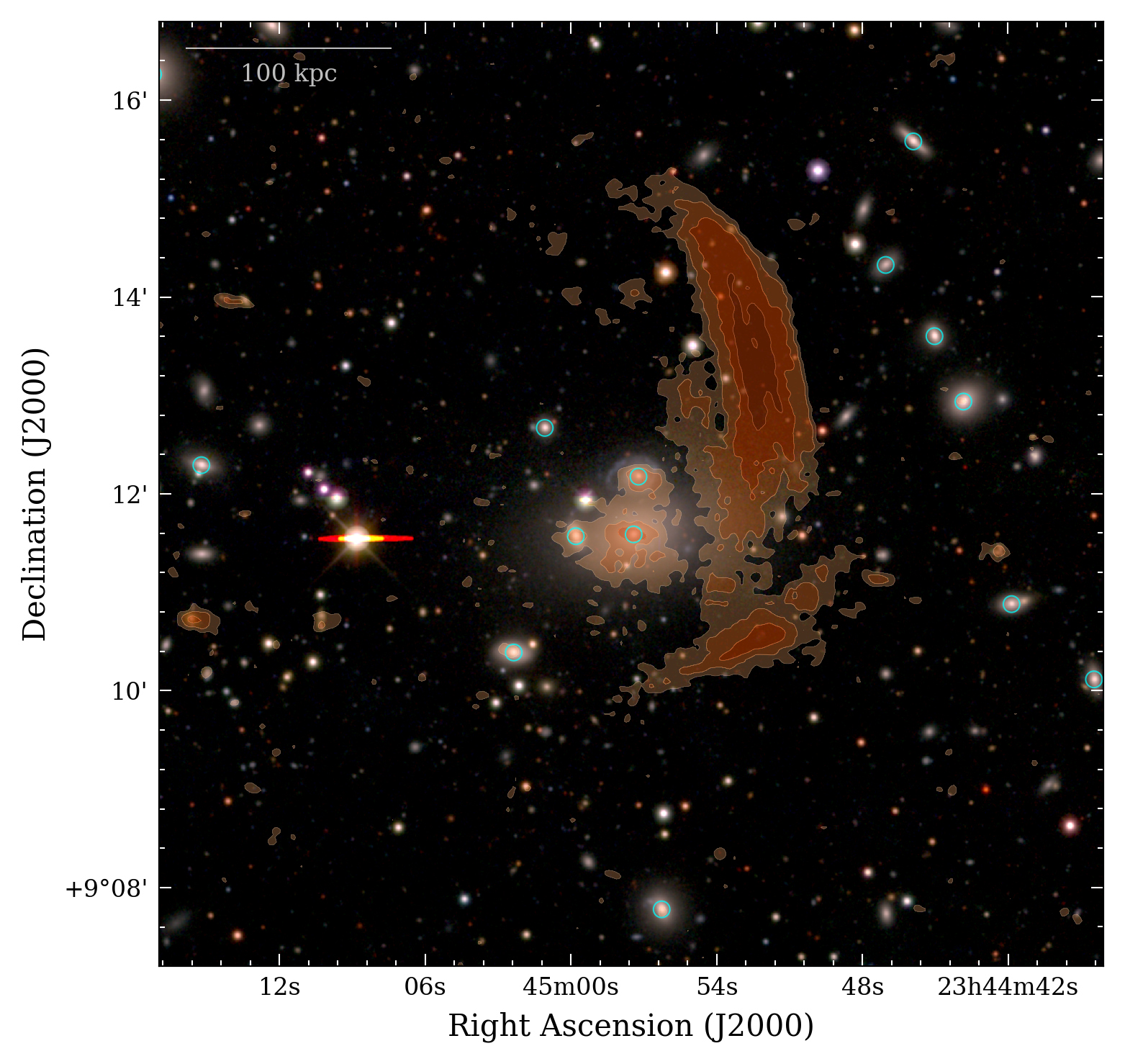}
  \caption{\textit{Left}: \lofar\ image at 144 MHz of the inner $\sim$450 kpc$^2$ of A2657 with the main features discussed in the text labeled. The noise of the image is $\sigma = 100$ \mujyb\ and the beam size, shown in the bottom left corner, is $13.2\arcsec \times 5.5\arcsec$. \textit{Right}: Composite optical/radio image of the same region. The optical \textit{g,r,z} image in the background is from the \desi\ \citep{dey19}. Orange colors denote the radio emission of the left panel, where contours are spaced by a factor of 2 from $3\sigma$. Cyan circles mark galaxies with redshift $0.035 < z < 0.045$ in the NASA/IPAC Extragalactic Database (NED).}
 \label{fig:composite}
\end{figure*}

\section{Data reduction}

\subsection{\lofar}

A2657 was observed in the context of the \lotssE\ \citep[\lotss;][]{shimwell17, shimwell19}. \lotss\ observations employ the \lofar\ High Band Antenna (HBA) with a 48 MHz bandwidth (120$-$168 MHz) centred at 144 MHz, reaching sensitivities $\sim$100 \mujyb\ at 6\arcsec\ resolution. These sensitivities are typically achieved with 8 or 12~hr duration observations depending on whether the survey pointing is at declination greater or less than 20\deg\ to mitigate the worse sensitivity of low-declination fields that are generally those observed at the lowest elevations. At the central frequency of HBA, the \lofar\ full width at half maximum is 3.96\deg. Since the pointings of the survey grid are separated by $\sim$2.6\deg, it occurred that A2657 was covered by two fields: P355+08 and P355+11, respectively centred at 0.97\deg\ and 1.85\deg\ from the target. These pointings do not belong to the latest publicly released sky area \citep[\ie\ \lotss-DR2;][]{shimwell22}, but will be part of a future data release. \\ 
\indent
\lotss\ observations are processed by the \lofar\ Surveys Key Science Project team by means of automated pipelines that produce direction-dependent corrected images of the survey pointings \citep{vanweeren16calibration, williams16,degasperin19, tasse21, shimwell22}. Image quality toward specific targets can be further improved by adopting the ``extraction+selfcal'' procedure described in \citet{vanweeren21}. In brief, this method consists of subtracting sources in the \uv-plane outside a small region containing the target of interest, phase-shifting the visibilities to the centre of this region, correcting for the \lofar\ primary beam toward this direction, and performing additional self-calibration loops. To date, this technique has been adopted for the analysis of hundreds of targets \citep[\eg][]{vanweeren21, botteon22dr2, heesen22}. We applied this strategy also to the two survey fields covering A2657, ``extracting'' a sky region of $33\arcmin\times33\arcmin$. The smaller field-of-view of the new data set also allows for fast re-imaging, which we performed with \wsclean\ v3.1.1 \citep{offringa14} enabling the multiscale multifrequency deconvolution \citep{offringa17} with a \texttt{robust=-0.5} weighting of the visibilities \citep{briggs95}, unless stated otherwise. \\
\indent
We searched for possible offsets in the \lofar\ flux density scale, which may occur because of inaccuracies in the \lofar\ beam model \citep[\eg][]{hardcastle16}, by cross-matching a catalogue of compact sources extracted from \lofar\ with the \nvssE\ \citep[\nvss;][]{condon98}, following the strategy described in \citet{hardcastle21} and \citet{shimwell22}. We found that our \lofar\ images needed to be corrected by a multiplicative factor of 1.30; throughout the paper, we adopt a conservative systematic uncertainty of 20\% on flux density measurements.

\subsection{\xmm}

We retrieved from the \xmm\ Science Archive the three observations available on A2657 (ObsIDs: 0300210601, 0402190301, 0505210301), with roughly the same exposure time, for a total of 91~ks. We processed the data using the \esasE\ \citep[\esas;][]{snowden08} embedded in the \xmm\ \sasE\ (\sas\ v16.1) following the analysis steps described in detail in \citet{ghirardini19universal}. We applied the soft protons cleaning procedure through the \texttt{mos-filter} and \texttt{pn-filter} tasks, employing light curves in the 2.5$-$8.5 keV energy range. For each camera on board \xmm\ (\ie\ MOS1, MOS2, pn), we generated photon-count images in the 0.7$-$1.2 keV band, a spectral range chosen to optimise the signal-to-noise of the cluster thermal emission \citep[\eg][]{ettori10}, and derived the corresponding exposure maps with the tool \texttt{eexpmap}. Following this, we implemented the strategy outlined in \citet{ghirardini19universal} to remove the particle background. This involved creating background images that accounted for both the particle background and residual soft protons. To maximise the statistic, we combined the photon-count images, exposure maps, and background files of the three cameras for the three processed ObsIDs. For the analysis of surface brightness profiles, we used \pyproffit\ \citep{eckert20}. \\
\indent
The spectral analysis has been performed by using \obsid\ 0300210601, since it was the only \obsid\ with sufficient exposure uncontaminated by soft protons ($\sim$10~ks) to perform a spectral study. We produced X-ray spectroscopic temperature maps following the wavelet filtering algorithm described in \citet{bourdin04, bourdin13, bourdin08}. This algorithm samples the field-of-view of the instrument by using square grids, with size increasing by powers of 2 and allowing at least 200 photons in each grid element. It extracts the spectrum and locally estimates the temperature $kT$ and its associated fluctuations $\sigma_{\rm kT}$ on various analysis scales by fitting a spectral model to the data. Then, the algorithm convolves the set of $kT$ and $\sigma_{\rm kT}$ by complementary high-pass and low-pass analysis filters to obtain wavelet coefficients. These wavelet coefficients are then subject to thresholding based on a 1$\sigma$ confidence level to reconstruct a denoised temperature map \citep[we refer the reader to][for further details]{bourdin04, bourdin08}.

\begin{figure}
 \centering
 \includegraphics[width=\hsize,trim={0.1cm 0.2cm 0.1cm 0.2cm},clip]{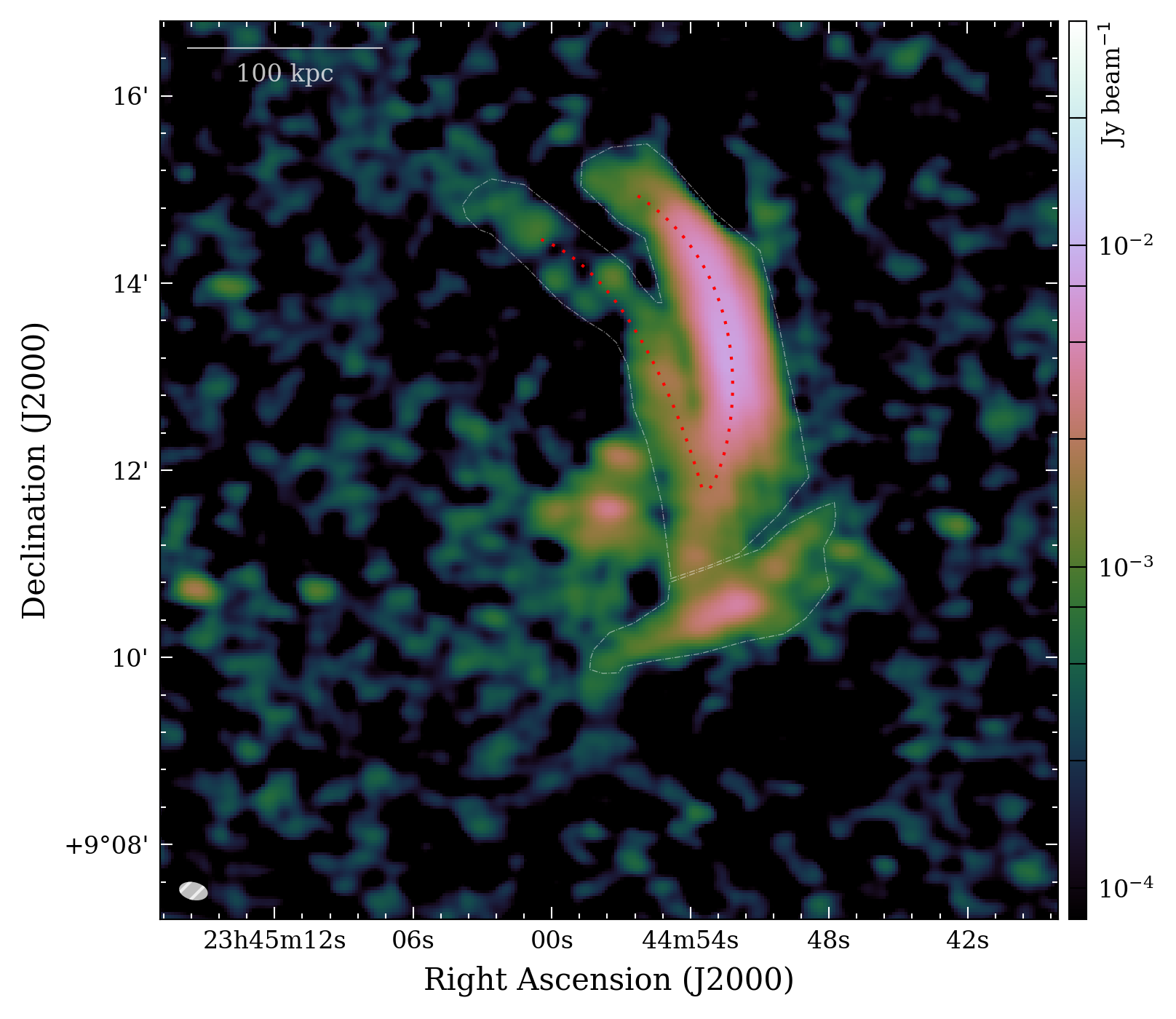}
  \caption{Low-resolution \lofar\ image at $18.7\arcsec \times 11.4\arcsec$ with noise $\sigma = 160$ \mujyb. The beam size is shown in the bottom left corner. Dotted red lines mark the bifurcated radio arc. Flux density measurements were performed in the regions delimited by dash-dotted white lines.}
 \label{fig:lowres}
\end{figure}

\begin{figure*}
 \centering
 \includegraphics[width=.49\hsize,trim={0cm 0cm 0cm 0cm},clip,valign=c]{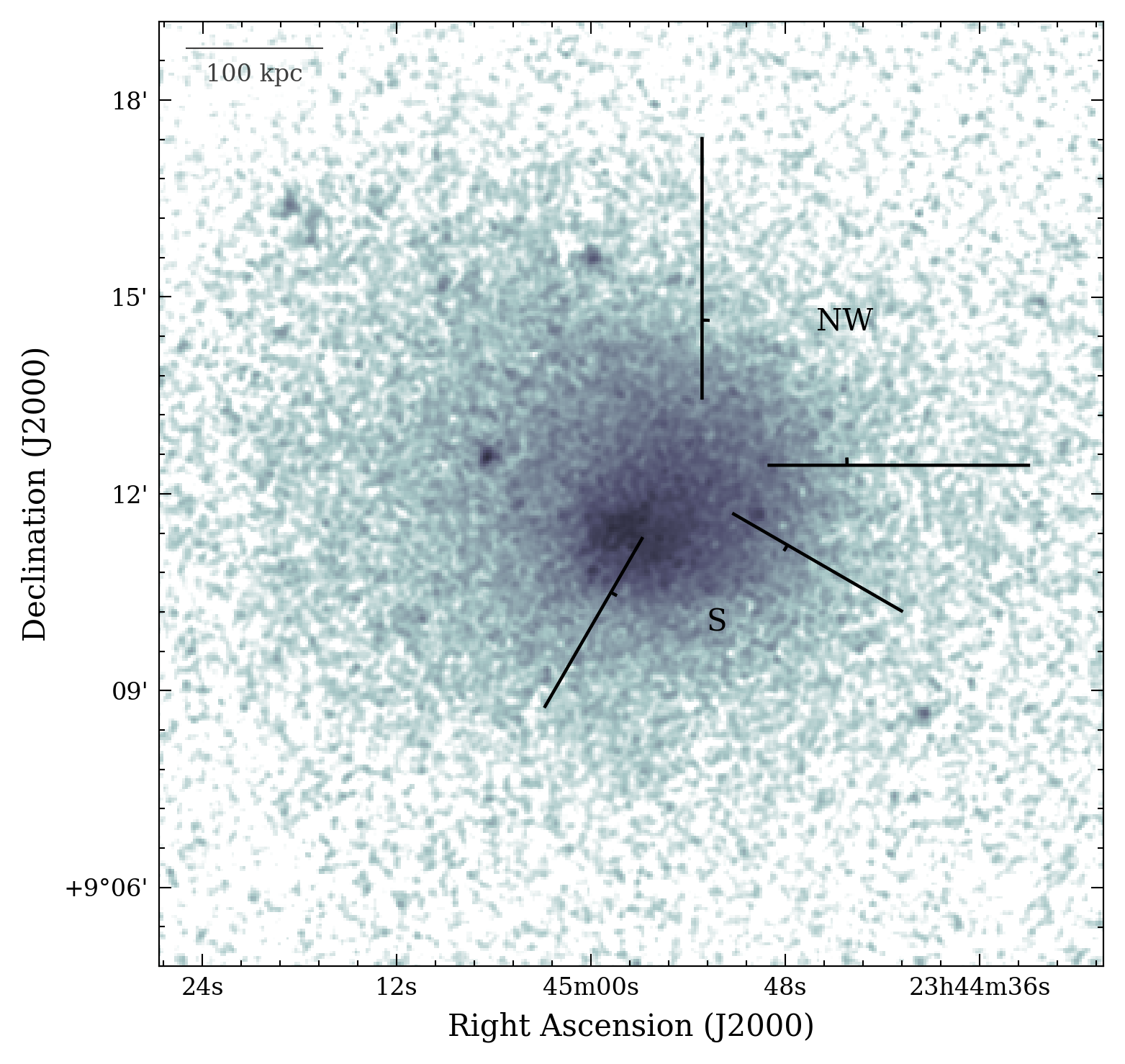}
 \includegraphics[width=.49\hsize,trim={0cm 0cm 0cm 0cm},clip,valign=c]{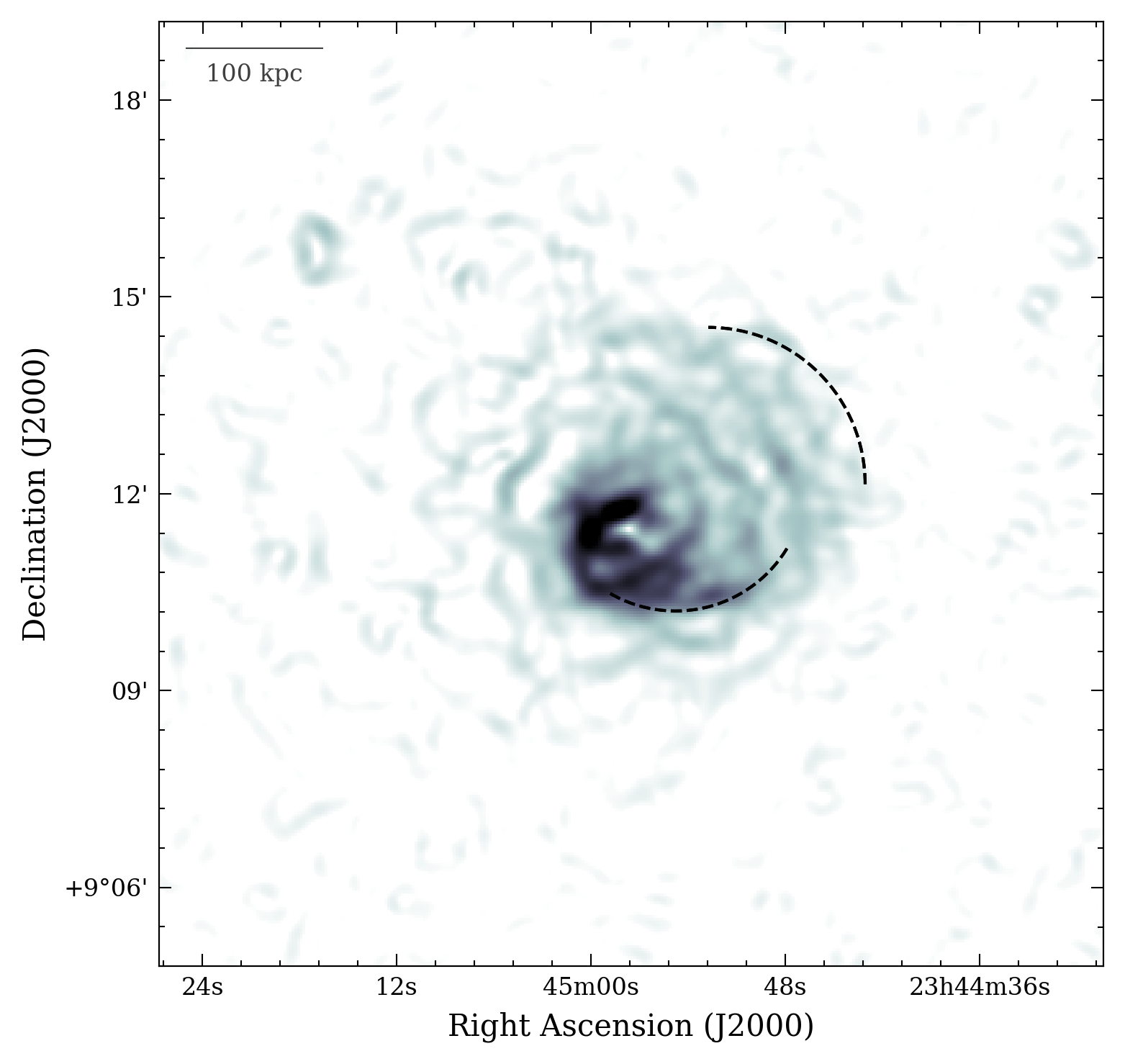}
 \includegraphics[width=.49\hsize,trim={0cm 0cm 0cm 0cm},clip,valign=c]{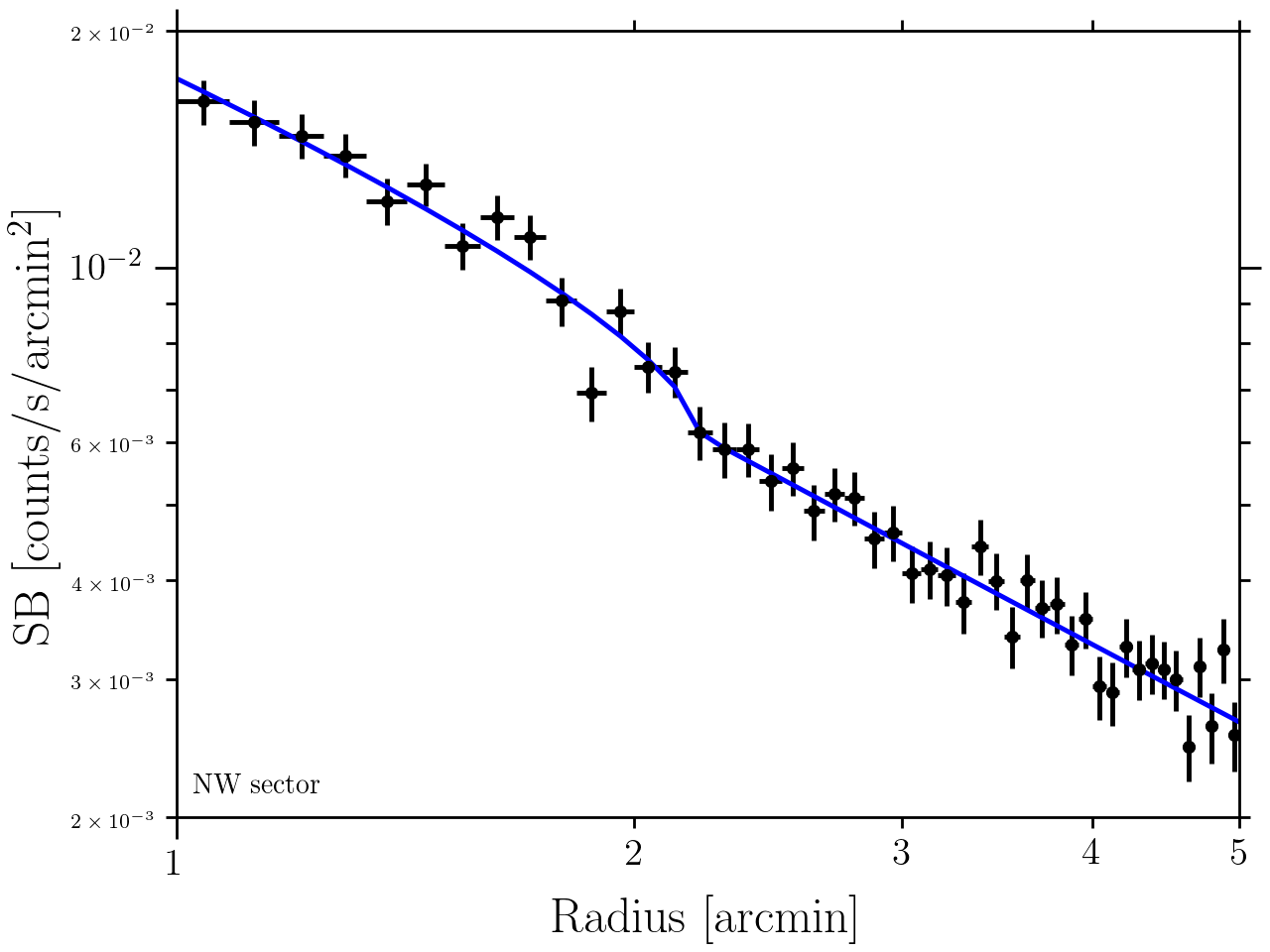}
 \includegraphics[width=.49\hsize,trim={0cm 0cm 0cm 0cm},clip,valign=c]{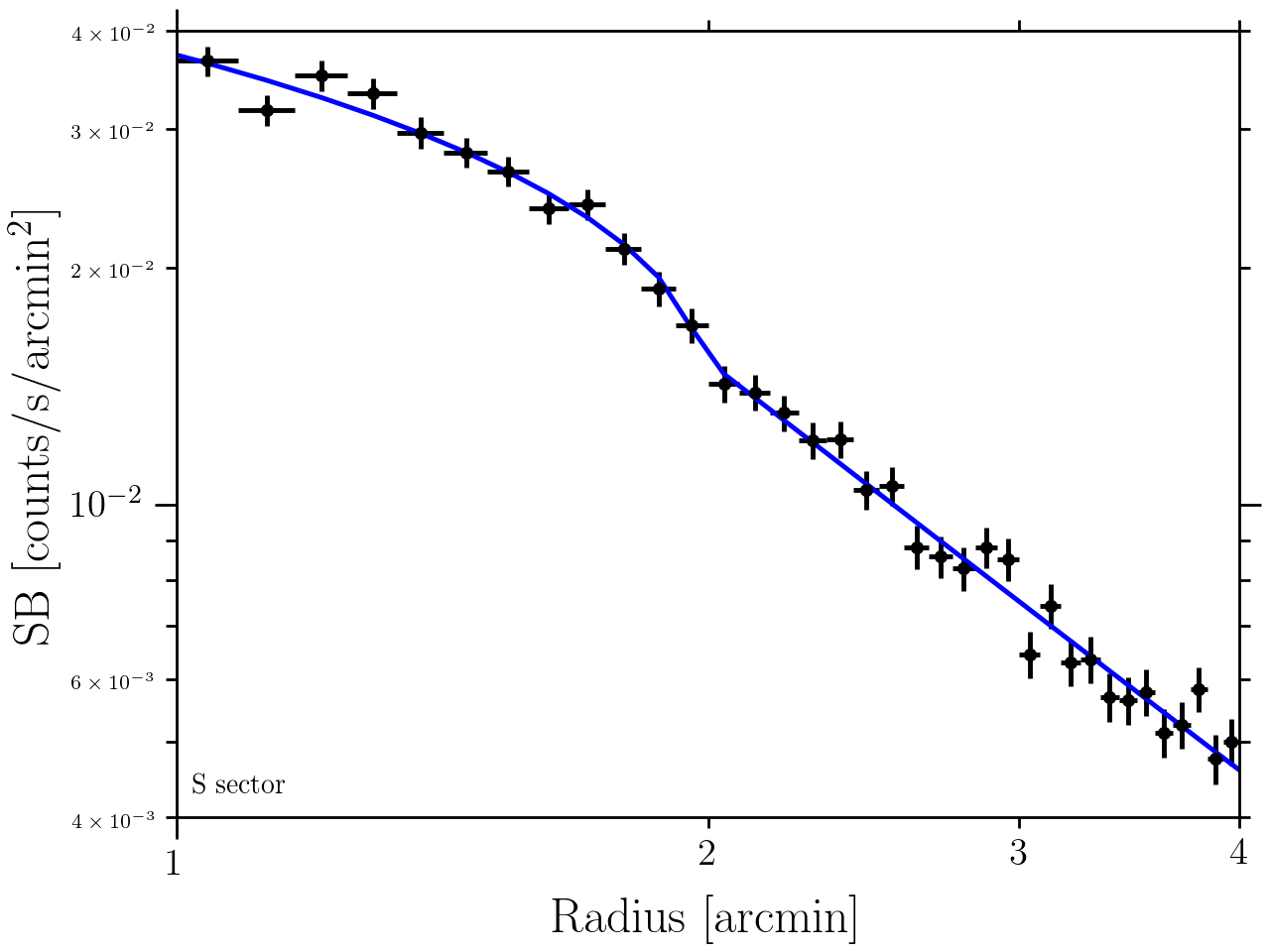} 
  \caption{\textit{Top panels:} Exposure-corrected background-subtracted \xmm\ image in the 0.7$-$1.2 keV band (\textit{left}) and corresponding GGM filtered image on scales of 10\arcsec\ (\textit{right}) of A2657. Sectors used for the analysis of surface brightness profiles are reported in black, with inner ticks/dashed arcs marking the best-fit surface brightness jump radii. \textit{Bottom panels:} Surface brightness profiles for the NW (\textit{left}) and S (\textit{right}) sectors with best-fit broken power-law model reported in blue. The $\chi^2$/d.o.f. of the fits are  $40.6/43$ (NW sector) and $33.2/31$ (S sector).}
 \label{fig:xmm}
\end{figure*}

\begin{figure*}
 \centering
 \includegraphics[width=\hsize,trim={0.2cm 0.4cm 0.2cm 0.2cm},clip]{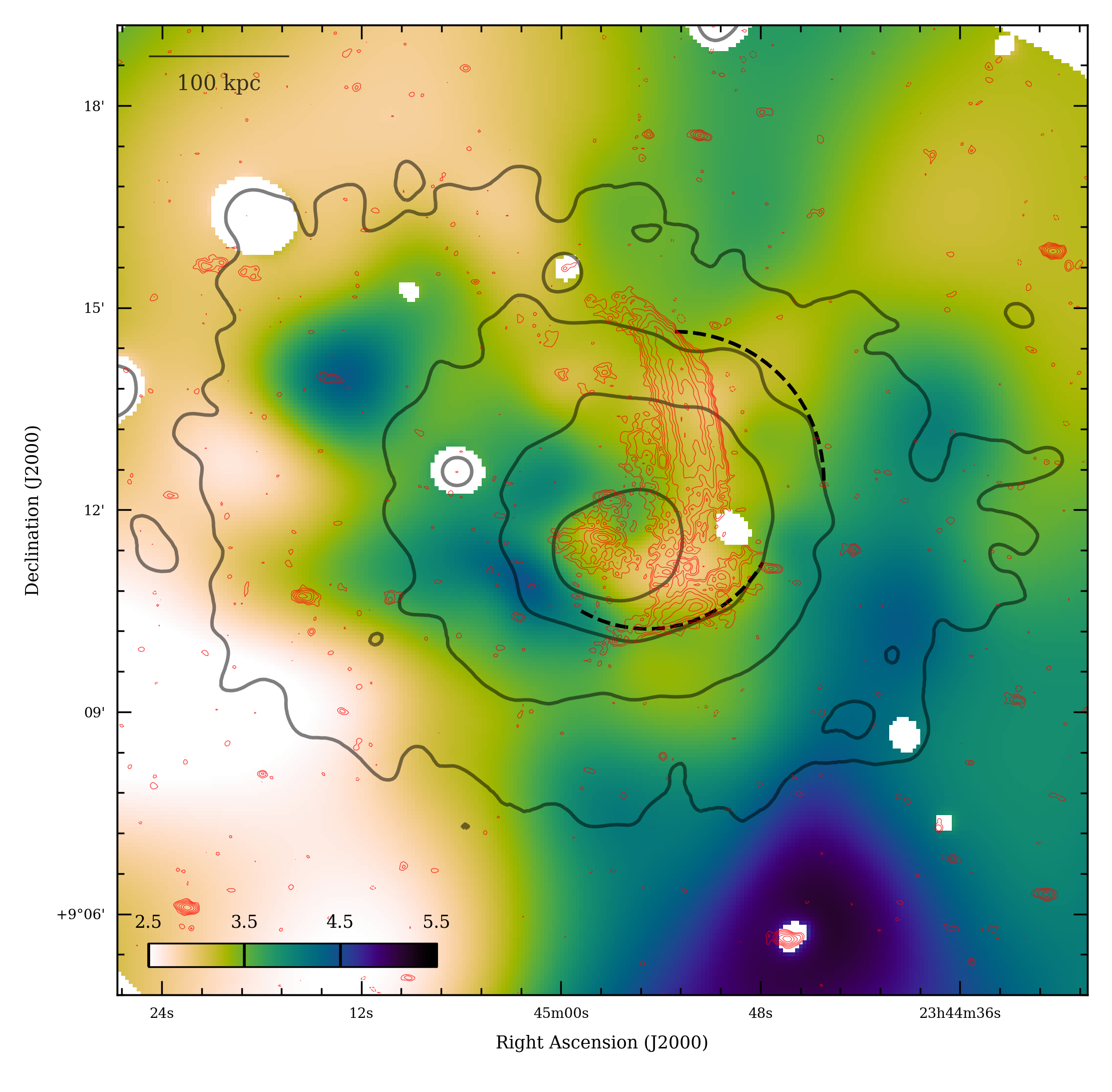}
  \caption{Temperature map in keV units obtained with \xmm. Contours denote the emission from \xmm\ (in \textit{thick black} lines) and from \lofar\ (in \textit{thin red} lines). \lofar\ contours are obtained from the image of Fig.~\ref{fig:composite}, and are spaced by a factor of $\sqrt{2}$ from $3\sigma$ (with the $-3\sigma$ contour reported in dashed). Dashed arcs are the same of Fig.~\ref{fig:xmm}. Blanked pixels correspond to regions masked out from the analysis because of the presence of point sources.}
 \label{fig:kt_map}
\end{figure*}

\begin{figure*}
 \centering
 \includegraphics[width=\hsize,trim={0.2cm 0.4cm 0.2cm 0.2cm},clip]{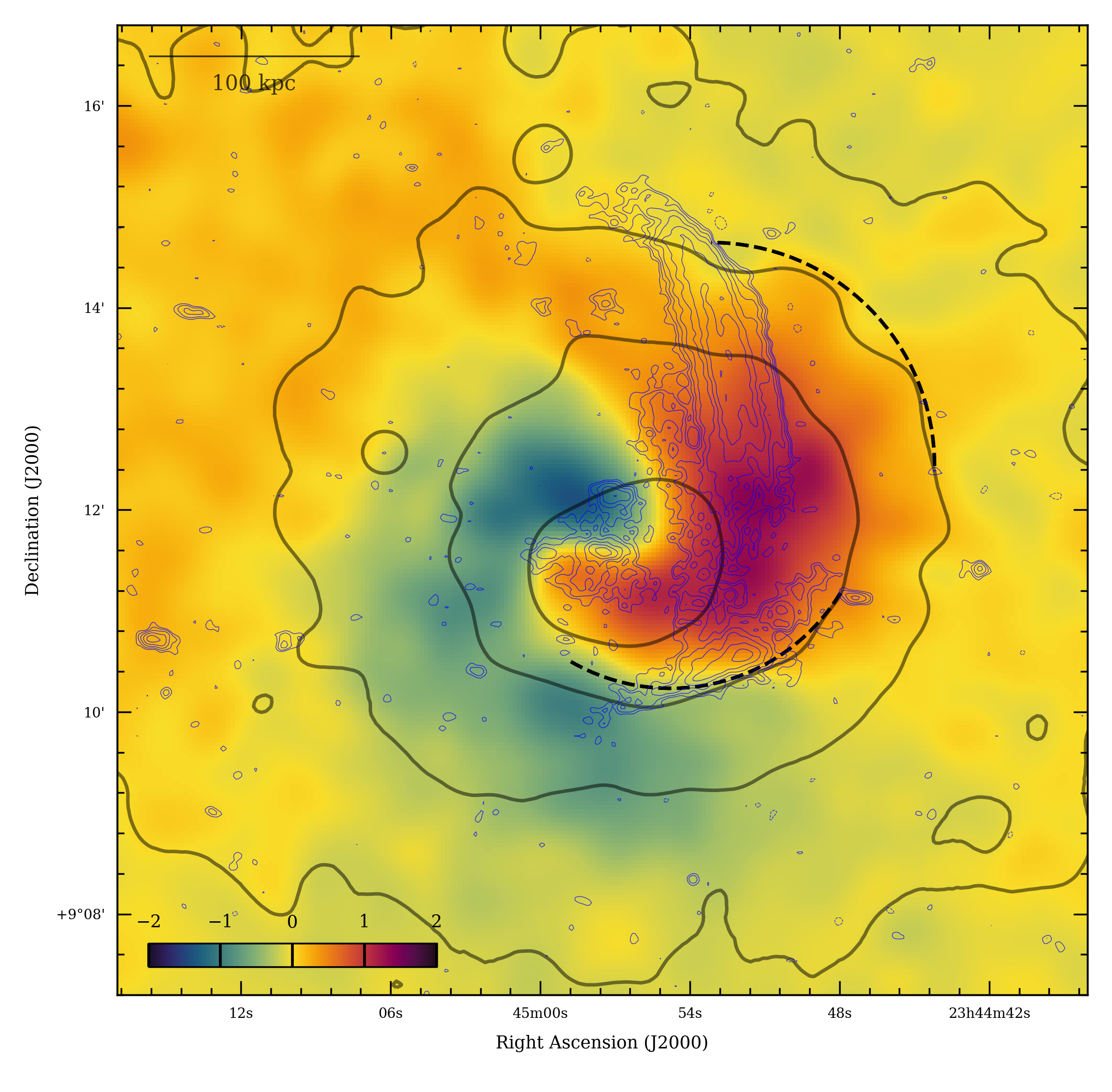}
  \caption{Residual X-ray emission obtained after subtracting the azimuthally averaged cluster emission from the \xmm\ 0.7$-$1.2 keV image. Contours and arcs are the same of Fig.~\ref{fig:kt_map}.}
 \label{fig:residuals}
\end{figure*}

\section{Results}

The \lofar\ image of A2657 of Fig.~\ref{fig:composite} shows the presence of a prominent diffuse arc-shaped emission at 144~MHz to the west of the BCG. The emission appears bifurcated into a brighter `outer' arc and a fainter `inner' arc, which blend together toward their southern edges. The radio emission between the two arcs is depleted, and the surface brightness dip increases in width south to north. We note that the northern region of the inner arc is detected at low-significance in Fig.~\ref{fig:composite}, where it appears patchy, but it is better recovered in the lower resolution image of Fig.~\ref{fig:lowres} (at the cost of a less defined separation between the two arcs), which was obtained with a 10\arcsec\ Gaussian taper on the visibilities. No clear optical counterparts can be identified for the two arcs nor for the patches of emission in the inner arc (\cf\ right-hand panel of Fig.~\ref{fig:composite}). The flux density at 144~MHz of the bifurcated arc measured in the white region reported in Fig.~\ref{fig:lowres} is $216\pm43$ mJy, $\sim$85\% of which is accounted by the outer arc. We note that the flux density error is dominated by the uncertainty on the \lofar\ flux scale, and that disentangling the contribution of each arc to the total measured flux density is somehow arbitrary as their emission blends in the southern direction. Based on the non-detection of diffuse emission in A2657 at 1.4~GHz from the \nvss\ \citep{condon98} and considering the mean surface brightness of the bifurcated radio arc at 144~MHz (8.06 $\mu$Jy arcsec$^2$), we estimate a $2\sigma$ ($3\sigma$) lower limit for its spectral index of $\alpha > 1.3$ ($\alpha > 1.1$). For the outer arc, the  $2\sigma$ ($3\sigma$) limit is $\alpha > 1.4$ ($\alpha > 1.2$). The radio emission extends for about 150$-$200~kpc in projection along the north-south direction. \\
\indent
Another diffuse, linear feature, that we labelled `F' in Fig.~\ref{fig:composite}, is observed with \lofar\ close to the centre of A2657. This radio emission is located $\sim$80~kpc south-west from the BCG, it is elongated for $\sim$125~kpc along its major axis, with a transverse size of $\sim$25~kpc (all distances are projected on the plane of the sky). Its flux density at 144~MHz measured in the white region reported in Fig.~\ref{fig:lowres} is $40\pm8$ mJy. In this case, the lower limit on the spectral index estimated from the source mean surface brightness at 144~MHz (5.55 $\mu$Jy arcsec$^2$) and the non-detection at 1.4~GHz from the \nvss\ at $2\sigma$ ($3\sigma$) is $\alpha > 1.1$ ($\alpha > 0.9$). \\
\indent
We summarise the main properties of the diffuse radio sources detected in A2657 in Tab.~\ref{tab:summary}. \\
\noindent
\begin{table}
 \centering
 \caption{Properties of the diffuse sources in A2657. Reported quantities are: largest-linear size (LLS), flux density ($S_{144}$) and radio power ($P_{144}$) at 144~MHz, and $2\sigma$ lower limit of the spectral index ($\alpha$).}\label{tab:summary}
 \begin{tabular}{lcccc} 
  \hline
  Source & LLS & $S_{144}$ & $P_{144}$ & $\alpha$ \\
         & (kpc) & (mJy) & ($\times 10^{23}$ \whz) & \\
  \hline
  Bifurcated arc & 150$-$200 & $216\pm43$ & $8.1 \pm 1.6$ & $>$1.3 \\
  F & 125 & $40\pm8$ & $1.5 \pm 0.3$ & $>$1.1 \\
  \hline
 \end{tabular}
\end{table}
\indent
In the \xmm\ 0.7$-$1.2~keV image and corresponding Gaussian Gradient Magnitude (GGM) filtered image of A2657 (top panels of Fig.~\ref{fig:xmm}), we observe signs of a mildly disturbed ICM. In particular, the GGM filter produces a smoothed image of the surface brightness gradient magnitude \citep[\eg][]{sanders16ggm}, and for the case of A2657 it highlights a spiral-like pattern of the thermal gas emission. We extract two surface brightness profiles across this spiral and report the presence of weak surface brightness discontinuities (bottom panels of Fig.~\ref{fig:xmm}) by fitting the data assuming an underlying density model consisting of two power-laws projected along the line-of-sight, $n_{\rm in} \propto r^{a}$ and $n_{\rm out} \propto r^{b}$, on the inner and outer sides of the edge, with a density jump \compr\ at the radial distance of the discontinuity $r_{\rm j}$. This is the typical model adopted to describe surface brightness edges in the ICM \citep[\eg][]{markevitch07rev}. The best-fit density and radial jumps for the northwest and south sectors are $\compr^{\rm NW} = 1.3\pm0.1$ and $\compr^{\rm S} = 1.3\pm0.1$, and $r_{\rm j}^{\rm NW} = 2.21^{+0.06}_{-0.07}$ and $r_{\rm j}^{\rm S} = 1.97^{+0.04}_{-0.09}$ arcmin. Quoted errors refer to the standard deviations associated with the fitting procedure. \\
\indent
The temperature distribution of the ICM can provide useful insights both on the dynamical state of the cluster and on the nature of the surface brightness discontinuities (as shocks and cold fronts show inverted temperature jumps). Our temperature map of A2657 shows a cold gas spiral extending in anticlockwise direction from the cluster centre (Fig.~\ref{fig:kt_map}). This gas spiral has a temperature of 2.9$-$3.2 keV and is mirrored in the residual image obtained by subtracting the azimuthally averaged cluster emission from the \xmm\ 0.7$-$1.2 keV image (Fig.~\ref{fig:residuals}). These are typical features of gas sloshing in the ICM \citep[\eg][]{ascasibar06}, and suggest that the surface brightness discontinuities detected in Fig.~\ref{fig:xmm} trace sloshing cold fronts. Interestingly, the bifurcated radio arc lays in projection on the dense and cold spiral and is roughly stretched along the same direction while source F is located at the interface of the sloshing gas spiral (Figs.~\ref{fig:kt_map} and \ref{fig:residuals}), where the southern surface brightness profile shows a discontinuity (bottom right panel of Fig.~\ref{fig:xmm}). As we shall discuss in Section~\ref{sec:origin}, this connection possibly gives a hint of the origin of the radio emission. 

\section{Discussion}

\subsection{Origin of the bifurcated radio arc}\label{sec:origin}

Diffuse non-thermal sources in the ICM are being observed in an increasing number of galaxy clusters \citep[\eg][for recent collections]{duchesne21eor, vanweeren21, botteon22dr2, hoang22, knowles22}. Despite this, we are not aware of any other system showing emission from a bifurcated arc such as that we detected in A2657. In dynamically relaxed sloshing clusters, it is possible to observe diffuse radio emission in the form of mini haloes \citep[\eg][]{giacintucci17}. However, these kinds of sources are typically roundish in morphology and fill the central cluster volume, with sizes roughly ranging from 50 to 500 kpc. Giant radio haloes span even larger scales ($>$1 Mpc) and are mostly detected in strongly disturbed clusters \citep[\eg][]{cassano10connection, cassano23}. Instead, arc-shaped emission is generally a distinctive characteristic of radio relics \citep[\eg][]{vanweeren10}, but in the case of relics the emission is offset with respect to the cluster central region, extends over large scales (even beyond the Mpc-scale), and the host cluster shows evidence of an ongoing major merger event. Neither of these classes of diffuse sources in the ICM seems to reflect the properties of the newly discovered emission in A2657. \\

\begin{figure*}
 \centering
 \includegraphics[width=.8\hsize,trim={0cm 0cm 0cm 0cm},clip]{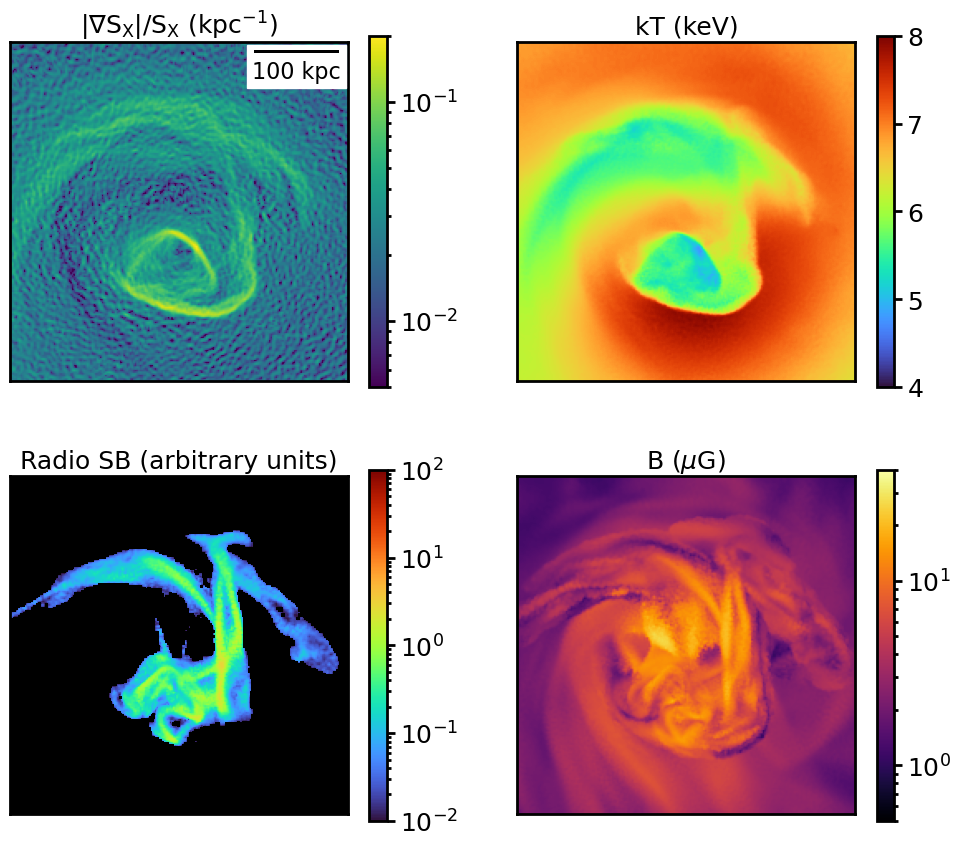}
  \caption{A snapshot from a simulation showing the X-ray surface brightness gradient computed using a GGM filter (\textit{top left}), the emission-weighted projected temperature (\textit{top right}), projected radio emission (\textit{bottom left}), and projected (mass-weighted) magnetic field (\textit{bottom right}) from an epoch chosen to resemble the bifurcated radio emission observed in A2657.}
 \label{fig:simulation}
\end{figure*}

\begin{figure*}
 \centering
 \includegraphics[width=\hsize,trim={0cm 0cm 0cm 0cm},clip]{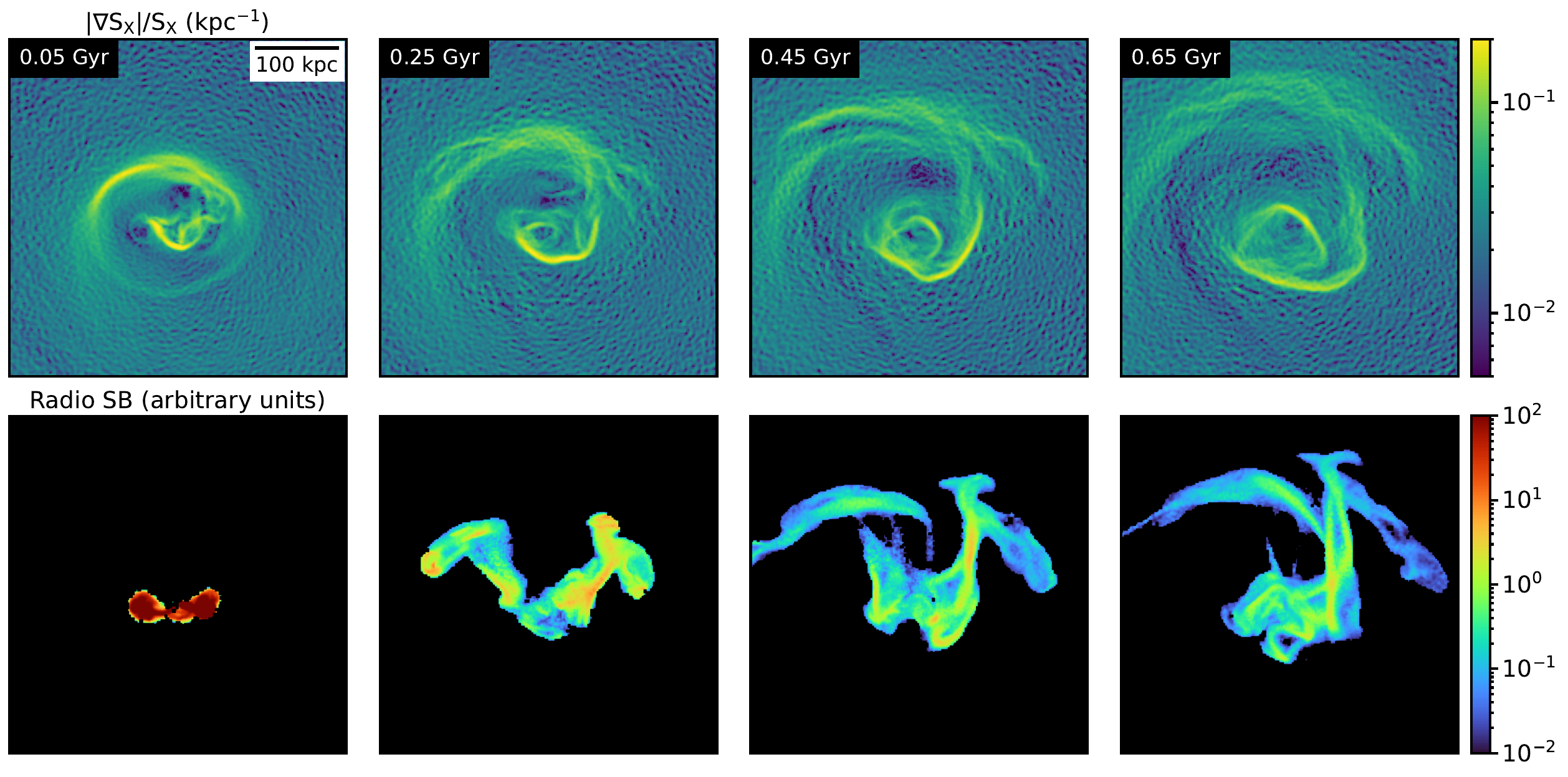}
  \caption{Evolution of the X-ray surface brightness gradient (\textit{top panels}) and radio emission (\textit{bottom panels}) from the simulation shown in Fig.~\ref{fig:simulation} at epochs $\Delta t = 0.05, 0.25, 0.45$ and 0.65 Gyr after the jet ignition.}
 \label{fig:epochs}
\end{figure*}

Although there is not an obvious connection between radio emission and cluster galaxies (Fig.~\ref{fig:composite}, right-hand panel), it is likely that the emission observed with \lofar\ at the centre of A2657 is related to AGN plasma. Low-frequency radio observations are indeed particularly efficient in the detection of ultra-steep radio emission due to fossil plasma ejected by past AGN activity. Eventually, this old plasma may be re-energised in the ICM via processes that are unrelated to the progenitor radio galaxy and that boost the emissivity of the aged radio plasma \citep[\eg][]{slee01, vanweeren09gmrt, cohen11, kale12relics, mandal20, hodgson21jellyfish}. These sources are sometimes referred to as radio phoenices, show amorphous morphologies, and a commonly adopted scenario to explain their origin is the adiabatic compression of old relativistic radio plasma due to the passage of a shock wave \citep[\eg][]{ensslin01, ensslin02relics}. Whilst a clear candidate to have ejected non-thermal components in A2657 could be its BCG (which is detected at 144 MHz and shows a diffuse component on scales of $\sim$45 kpc surrounding its nucleus), the presence of shocks capable of adiabatically compressing the plasma on scales $>$100~kpc is more difficult to explain in such a dynamically relaxed system (weak shocks from AGN outbursts typically span a few tens of kpc, see \eg\ \citealt{mcnamara12rev} and references therein). Nonetheless, the relativistic plasma may also interact with the gas which is sloshing in the cluster gravitational potential well. Radio structures similar to source F in A2657 have indeed been pointed out by recent \lofar\ observations of other sloshing systems (\eg\ A1775, \citealt{botteon21a1775}; NGC507, \citealt{brienza22}), where steep spectrum radio emission at the interface of the cold gas spiral has been interpreted as remnant plasma advected, compressed, and re-energised by the sloshing motion. Strong shear motions under cold fronts have also been suggested to be responsible of the abrupt deflections observed in some radio galaxy tails (\eg\ NGC1272 in the Perseus cluster, \citealt{gendronmarsolais21}; J1333-3141 in A3562, \citealt{giacintucci22}; PKS 0446-20 in A514, \citealt{lee23arx}). Moreover, patches of diffuse radio emission may also be generated due to the re-acceleration of relativistic plasma as a consequence of the intermittent and patchy turbulent motions in the ICM induced by gas sloshing \citep[\eg][]{zuhone13}. The bifurcated radio arc is instead a more regular structure with a very peculiar morphology, and in the following we propose a possible origin for this feature aided by results from recent MHD simulations with passive tracers devoted to the study of the evolution of non-thermal plasma in sloshing clusters. \\
\indent
In \citet{zuhone21transport, zuhone21bubbles}, MHD simulations were used to explore the possibility that AGN-blown radio lobes could be turned into radio relics by sloshing gas motions. They tested two types of off-axis binary cluster mergers that differ only in the properties of the sub cluster. They also adopted two different approaches to inject relativistic electrons in the ICM: bidirectional jets from the central AGN and isolated bubble from an off-centre radio galaxy. In all cases, the non-thermal plasma is found to swirl in spiral motions, following the thermal gas. This leads to the generation of structures that are stretched due to strong tangential motions induced by sloshing. In particular, the distribution of the passive tracers of a single bubble injected in the less violent merger configuration (which produces gentle and prolonged sloshing motions) is shown to take a bifurcated morphology under a certain viewing angle \citep[see Fig.~11 in][]{zuhone21bubbles}. A single bubble that is extended over $\sim$a few tens of kpc will encounter different ICM conditions at different locations on its surface. The result is that the bubble breaks apart as different parts of it mix with the surrounding ICM to different degrees (acquiring different entropies and thus bouyancies), and acquire different velocities. This feature in the simulations resembles the morphology of the radio emission observed in our \lofar\ images of A2657. We note that in these simulations the arcs extend for $>$1~Mpc, which is greater than the size of the radio emission detected in A2657. However, in these simulations, the radio bubble was placed at a large distance (200~kpc) from the cluster centre to test whether gas motions could spread the passive tracers for Mpc-scales, which are the typical extents of radio relics. \\
\indent
More recently, \citet{fabian22interaction} investigated the evolution of radio bubbles in cool-core clusters with MHD simulations of a Perseus-like system. They showed that bubbles that are buoyantly raising in a sloshing ICM and interact with the sloshing cold fronts can be dislocated and disrupted by thermal gas motions. In particular, bubbles can acquire elongated morphologies due to the tangential flows associated with the sloshing \citep[see Appendix A in][]{fabian22interaction}. In that paper, snapshots from simulations are reported until 150~Myr after the core passage of the sub cluster. As gas motions persist, the bubbles can be expected to become increasingly distorted and stretched. For this reason, we produced additional snapshots at later stages of the sloshing/bubbles interaction from the same simulations of \citet{fabian22interaction}. Details of the simulations are given in that work, but we highlight their most important characteristics here. \\
\indent
The simulations use the \textsc{arepo} code \citep{springel10,pakmor13,marinacci18} to evolve the equations of MHD, dark matter, and self-gravity. The simulations model a cool-core cluster with $M \sim 10^{15}$ \msun, with initial conditions taken from previous works \citep{ascasibar06, zuhone10, zuhone16cores, zuhone18hitomi, zuhone19sloshing, zuhone21bubbles}. This particular cluster experiences an off-axis encounter with a sub cluster of $M \sim 2 \times 10^{14}~M_\odot$ with only dark matter. The merger plane of this sub cluster lies entirely within the $x$--$y$ plane of the simulation domain, initiating sloshing motions that lead to the creation of cold fronts. The ICM in the simulation has an initially tangled magnetic field with an initial plasma parameter $\beta = p_{\rm th}/p_{\rm B} = 100$ (where $p_{\rm th}$ and $p_{\rm B}$ are the thermal and magnetic pressure, respectively). To model the effect of AGN jets, we use the method of \citet{weinberger17}, which injects a bidirectional jet that is kinetically dominated, low density, and collimated. Kinetic, thermal, and magnetic energy is injected into two small spherical regions situated a few kpc from the location of a black hole particle positioned at the cluster potential minimum. The material injected by the jet consists of a 50/50 mixture of two fluids. One fluid possesses an adiabatic index of $\gamma = 4/3$, representing cosmic rays, while the other has an adiabatic index of $\gamma = 5/3$, the same as the ICM. The cosmic ray fluid adheres to the same momentum and energy equations as the ICM, except for the difference in adiabatic index. In the simulation shown here, the jets are fired in the $x$-direction with a power $P_{\rm jet} = 10^{45}$~erg~s$^{-1}$ for a duration of $t_{\rm jet} = 50$~Myr, so the resulting total energy injected is $E_{\rm jet} \sim 1.6 \times 10^{60}$~erg in each direction, which is a sum of kinetic, thermal, and magnetic energy. Within the jet region, the magnetic and thermal pressures are balanced ($\beta_{\rm jet} = 1$), and the injected magnetic field is purely toroidal. The jets are injected 0.75~Gyr after the closest passage of the sub cluster. \\
\begin{figure*}
 \centering
 \includegraphics[width=.43\hsize,trim={0.1cm 0.2cm 0.1cm 0.2cm},clip,valign=c]{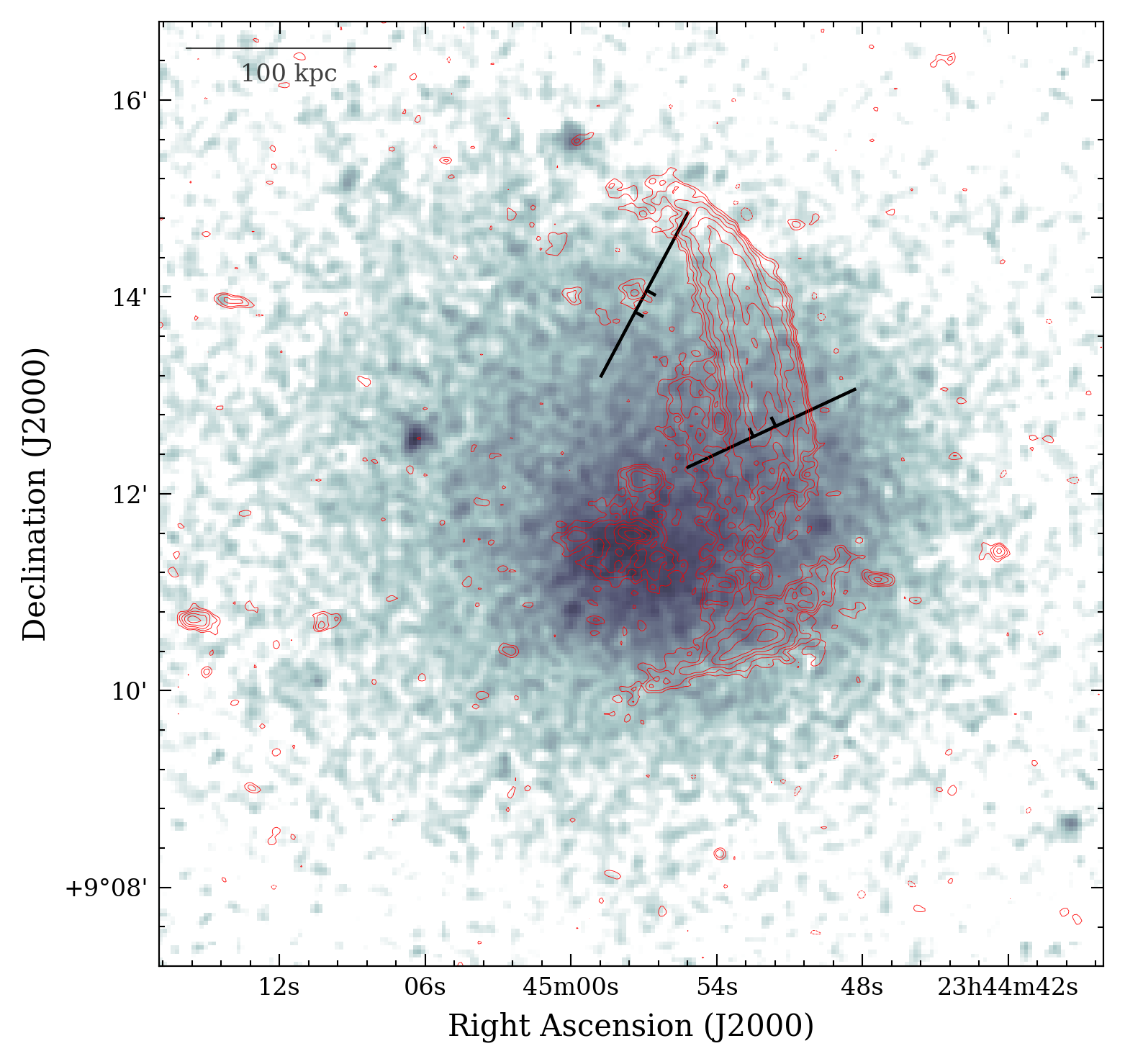}
 \hspace{1cm}
 \includegraphics[width=.45\hsize,trim={0cm 0cm 0cm 0cm},clip,valign=c]{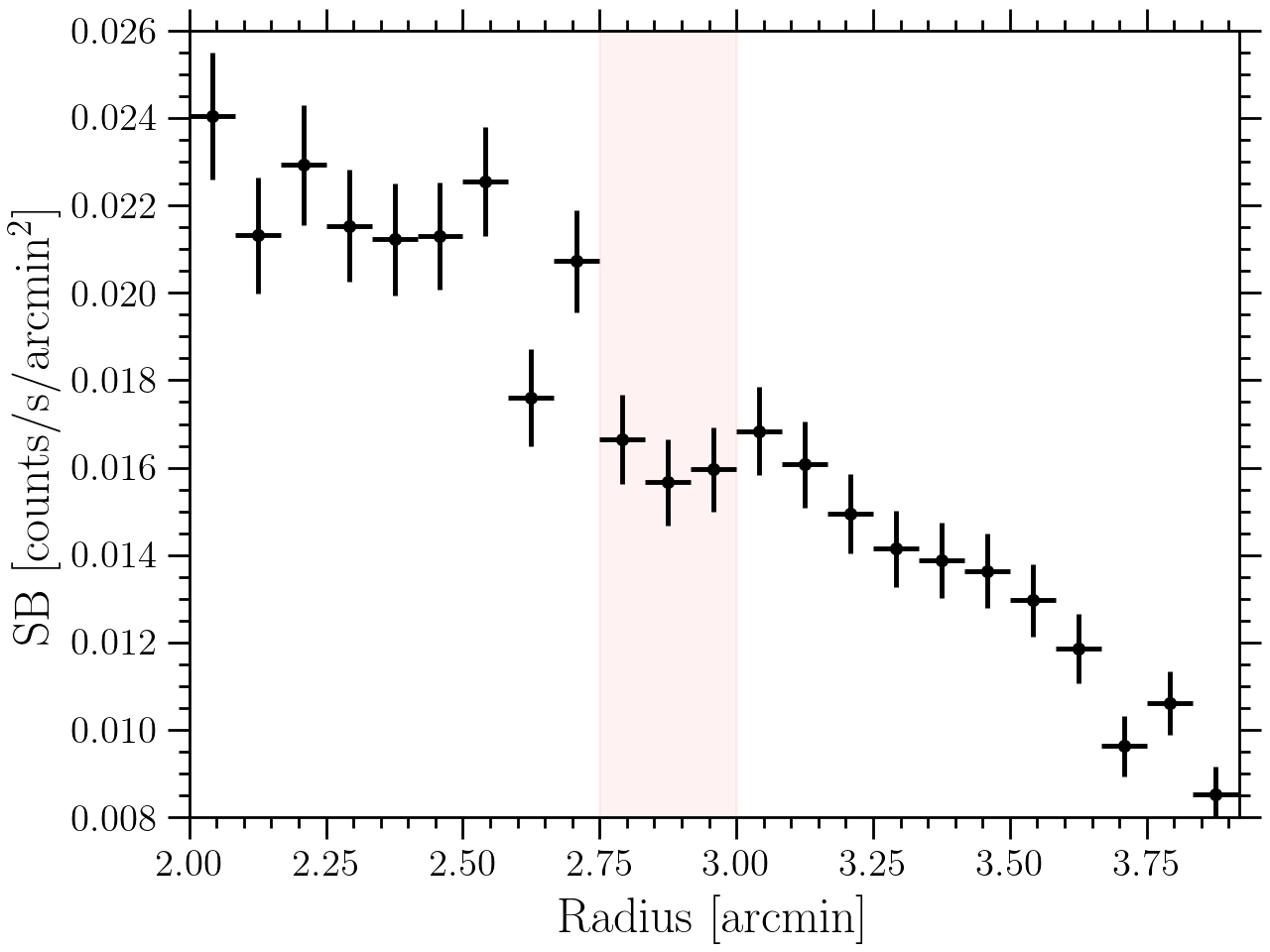}
  \caption{In the \xmm\ image on the \textit{left} panel we report the sector used to extract the X-ray surface brightness profile shown in the \textit{right} panel. Inner ticks in the sector mark the radial range highlighted in the profile that may exhibit a surface brightness dip. Radio contours are the same of Fig.~\ref{fig:kt_map}.}
 \label{fig:xray_dip}
\end{figure*}
\indent
In Fig.~\ref{fig:simulation} we report new snapshots of the simulated cluster core region for a projection and epoch selected to closely match the case of A2657. In Fig.~\ref{fig:epochs} we instead show snapshots at different epochs (measured from the moment of jet injection) showing the evolution of the X-ray and radio emission from the initial inflation of the bubbles to the end condition. The comparison of our observations with the MHD simulations discussed above gives a tantalising indication that A2657 may represent a case where an AGN-injected radio bubble evacuated by an outflow of the BCG is being shredded as a consequence of the interplay with the ongoing sloshing gas. We note that tangential flows can stir and compress non-thermal components in the ICM, where synchrotron radio emission can arise due to the enhanced density of relativistic electrons and the formation of strongly magnetised layers \citep[\eg][]{zuhone11}. However, when the radiative age of the emitting plasma is shorter than the magnetic field amplification timescale, \ie\ $t_{\rm rad} < t_{\rm B}$, the cooling of relativistic electrons becomes shorter and the bulk of synchrotron radiation is progressively emitted at very low frequencies. This scenario is in line with our simulations that suggest $t_{\rm B}$ of the same order of the timescale necessary for a bubble to be shredded by sloshing to produce an emission stretched for $>$100 kpc, i.e. $\sim$500 Myr (\cf\ Fig.~\ref{fig:epochs}), compared to $t_{\rm rad} \sim 115$ Myr (considering 144 MHz and a reference magnetic field $B = 10$ \muG). The fact that $t_{\rm rad}$ is shorter than both $t_{\rm B}$ and the time necessary to transport electrons on $>$100 kpc scale may require the presence of re-acceleration mechanisms to keep electrons radiating on the observed large scale structure detected with \lofar\ in A2657 (see \citealt{rudnick22} for the case of 3C40). In this respect, future radio observations will be very important to measure the integrated and resolved spectral properties of the emission and constrain its origin. \\ 
\indent
One of the most striking pieces of evidence of the interplay between thermal and non-thermal components in the ICM is provided by the connection between substructures observed in X-rays that are mirrored in radio and vice versa \citep[\eg][]{botteon23}. From a closer inspection of X-ray images of A2657, we noticed the possible presence of underdense gas in the region between the inner and outer arcs. For this reason, we extracted a surface brightness profile from the \xmm\ image across this region. After experimenting with sectors with different apertures, radial ranges, and centres, we adopted the sector shown in Fig.~\ref{fig:xray_dip}, which led to the surface brightness profile reported in the same figure. This sector was chosen to maximise what may be a depletion of the X-ray emission co-located with the region where the two radio arcs split. We note that the significance of this dip in the X-ray surface brightness profile depends on the sector choice possibly because of its subtle nature and the large PSF of \xmm. Unfortunately, the only existing observation of A2657 in the \chandra\ archive is too shallow to investigate this possible feature at higher angular resolution. If confirmed by deeper observations, this could represent an interesting case where there is a co-spatial bifurcation/split of the radio and X-ray emission in the ICM. We note that depletions of the X-ray emission have been reported in other galaxy clusters. These may trace plasma depletion layers where the thermal gas is squeezed out of narrow strongly magnetised layers because of the enhanced magnetic pressure therein \citep[\eg][]{wang16a520, wang18a2142} or hydrodynamical features that develop due to instabilities in the flow that produce bay- and hook-like structures \citep[\eg][]{roediger11, roediger12a496, zuhone18catalog, walker18split, walker22}. As the radio emission in A2657 is depleted in the region of the possible X-ray surface brightness dip, it is unlikely that this feature is related to a strongly magnetised layer, where the radio emission should be instead enhanced. Additionally, an X-ray surface brightness decrement tangential to the cluster centre may also be a consequence of the stretching and flattening of a cavity that is rising in the ICM while approaching a sloshing cold front surface \citep{fabian22interaction}.

\subsection{Thin strands in the outer arc}

\begin{figure}
 \centering
 \includegraphics[width=\hsize,trim={1.2cm 1.2cm 1.2cm 0.8cm},clip]{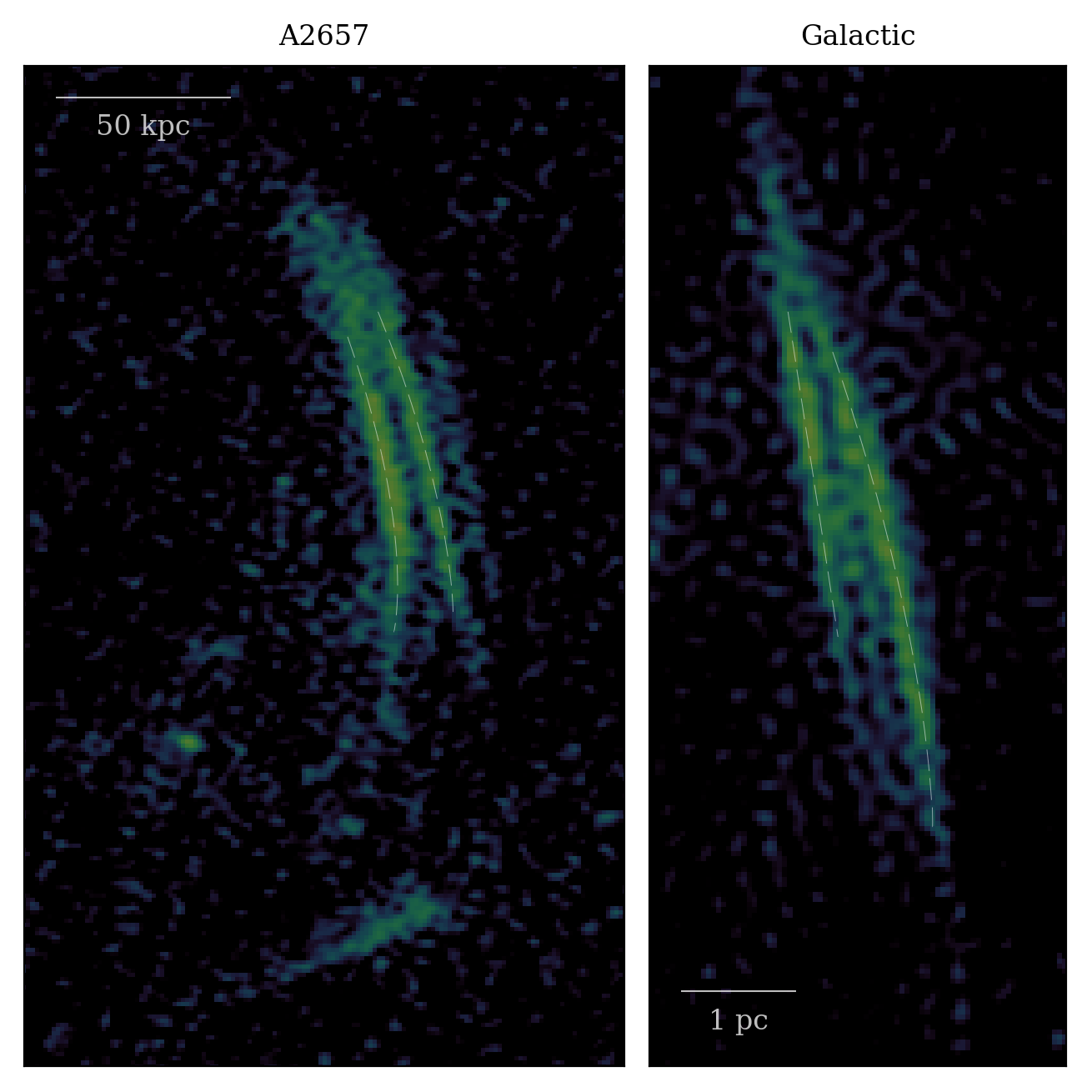}
  \caption{Comparison between the thin filaments of the outer arc in A2657 (\textit{left}) and the thin filaments observed in the Galactic plane (\textit{right}). The \textit{left} panel shows a \lofar\ image of A2657 at 144~MHz ($6.2\arcsec \times 4.0\arcsec$) while the \textit{right} panel shows a zoom-in of the \meerkat\ Galactic centre mosaic at 1.28~GHz ($4.0\arcsec \times 4.0\arcsec$) by \citet{heywood22galacticcenter}.}
 \label{fig:filaments}
\end{figure}

In the \lofar\ image of Fig.~\ref{fig:composite}, we noticed the presence of two parallel strands of emission embedded in the outer arc, stretched along its major axis. To better highlight these strands, we produced a higher resolution image with \texttt{robust=-1.25}, which we show in the left-hand panel of Fig.~\ref{fig:filaments}. These structures are resolved with a width of 6$-$7~kpc and extend for $\sim$150~kpc. \\
\indent
Non-thermal filaments are being observed in an increasing number of synchrotron sources. Remarkable cases where thin and parallel strands of emission were observed include: the radio relics in the Sausage \citep{digennaro18sausage} and Toothbrush \citep{rajpurohit18, rajpurohit20toothbrush} clusters; the cluster/group radio galaxies ESO137-006 \citep{ramatsoku20eso137}, Trail/T-bone/Original TRG complex in A2255 \citep{botteon20a2255, botteon22a2255}, IC4296 \citep{condon21}, Nest200047 \citep{brienza21}, and 3C40 \citep{rudnick22}; and the large population of filaments in the Galactic centre \citep[see][for recent work]{yusefzadeh22b, yusefzadeh22c, yusefzadeh22a}. Recently, \citet{yusefzadeh22filaments} pointed out the morphological similarity between the magnetised filaments in the ICM and Galactic centre, arguing that they may originate from similar processes occurring on very different scales. In Fig.~\ref{fig:filaments}, we also found an analogy between the strands detected in the outer arc in A2657 and one of the filaments observed in the Galactic centre (namely, G359.563+0.249, nicknamed the ‘Edge-on Spiral’ by \citealt{yusefzadeh22b}). The origin of these structures is still under investigation and their study could unveil important aspects of the microphysics of weakly magnetised plasma, such as the particle acceleration mechanisms, the amplification of the magnetic field, and the propagation mode of the relativistic electrons therein \citep{thomas20, rudnick22, yusefzadeh22filaments, ruszkowski23arx}.

\section{Conclusions}

We reported on the detection of peculiar diffuse radio emission in the central region of the nearby galaxy cluster A2657. The emission was discovered thanks to \lofar\ observations at 144~MHz, and consists of a bifurcated radio arc and of an additional diffuse feature that we labelled F. Based on the non-detection of diffuse emission in A2657 at 1.4~GHz from the \nvss, we estimated $2\sigma$ limits on the spectral index of $\alpha > 1.3$ and $\alpha > 1.1$ for the bifurcated arc and source F, respectively. From the analysis of archival \xmm\ observations, we found clear signatures of gas sloshing, demonstrated by a spiral of cold gas in the temperature map which is mirrored in the GGM-filtered and X-ray residual images, and by the presence of cold fronts. The bifurcated radio arc is stretched along the gas spiral while source F is located at the interface of the cold gas. This likely suggests a connection between the origin of the non-thermal emission and gas motions in the ICM. \\
\indent
We searched for structures analogous to those observed in A2657 in recent MHD simulations \citep{zuhone21transport, zuhone21bubbles, fabian22interaction} that investigate the evolution of non-thermal components in sloshing clusters. We found that relativistic electrons and magnetic fields can take similar elongated and bifurcated morphologies as a consequence of the strong tangential flows induced by the spiral motion of the thermal gas. We thus speculated that the bifurcated radio emission in the centre of A2657 traces an AGN-bubble that was injected into the ICM and later shredded by gas sloshing. \\
\indent
A tentative dip in the X-ray surface brightness profile across the two radio arcs is reported with \xmm. The dip is co-spatial with the region where the bifurcated radio emission depletes. A deep X-ray follow-up observation with \chandra\ would allow to firmly claim the dip of X-ray emission and study its properties and connection with the non-thermal plasma. We also report on the presence of two thin parallel strands of radio emission embedded in the outer arc similar to those observed in other cluster diffuse sources, radio galaxies, and in the Galactic centre. In this case, follow-up radio observations would be desired to study the spectral trends of the non-thermal emission, which would be beneficial both for understanding the origin of the thin strands and corroborating the proposed scenario of radio bubble shredded by gas sloshing. As tangential motions can stir magnetic fields and these features stretch along the sloshing spiral, at high frequency we may expect to detect polarised emission due to the presence of aligned magnetic fields. \\
\indent
This work shows the potential of the joint analysis of radio/X-ray observations of galaxy clusters supported by the interpretation with MHD numerical simulations. The combination of observational and numerical results is powerful for the investigation and understanding of the strong interplay between thermal and non-thermal components in the ICM.

\section*{Acknowledgements}
We thank the anonymous referee for the timely report and comments. ABotteon acknowledges financial support from the European Union - Next Generation EU.
JAZ acknowledges support from the \chandra\ X-ray Center, which is operated by the Smithsonian Astrophysical Observatory for and on behalf of NASA under contract NAS8-03060. 
ABonafede acknowledges support from ERC-Stg DRANOEL n. 714245 and MIUR FARE grant ``SMS''.
RJvW acknowledges support from the ERC-Stg ClusterWeb 804208. 
LOFAR \citep{vanhaarlem13} is the LOw Frequency ARray designed and constructed by ASTRON. It has observing, data processing, and data storage facilities in several countries, which are owned by various parties (each with their own funding sources), and are collectively operated by the ILT foundation under a joint scientific policy. The ILT resources have benefitted from the following recent major funding sources: CNRS-INSU, Observatoire de Paris and Universit\'{e} d'Orl\'{e}ans, France; BMBF, MIWF-NRW, MPG, Germany; Science Foundation Ireland (SFI), Department of Business, Enterprise and Innovation (DBEI), Ireland; NWO, The Netherlands; The Science and Technology Facilities Council, UK; Ministry of Science and Higher Education, Poland; Istituto Nazionale di Astrofisica (INAF), Italy. This research made use of the Dutch national e-infrastructure with support of the SURF Cooperative (e-infra 180169) and the LOFAR e-infra group, and of the LOFAR-IT computing infrastructure supported and operated by INAF, and by the Physics Dept.~of Turin University (under the agreement with Consorzio Interuniversitario per la Fisica Spaziale) at the C3S Supercomputing Centre, Italy. The J\"{u}lich LOFAR Long Term Archive and the German LOFAR network are both coordinated and operated by the J\"{u}lich Supercomputing Centre (JSC), and computing resources on the supercomputer JUWELS at JSC were provided by the Gauss Centre for Supercomputing e.V. (grant CHTB00) through the John von Neumann Institute for Computing (NIC). This research made use of the University of Hertfordshire high-performance computing facility and the LOFAR-UK computing facility located at the University of Hertfordshire and supported by STFC [ST/P000096/1].
The simulations were performed on the ``Pleiades'' high-performance computing system at the NASA Advanced Supercomputing facility at NASA/Ames Research Center.
This research made use of the NASA/IPAC Extragalactic Database (NED), operated by the Jet Propulsion Laboratory (California Institute of Technology), under contract with the National Aeronautics and Space Administration.
The scientific results reported in this article are based in part on observations obtained with \xmm, an ESA science mission with instruments and contributions directly funded by ESA Member States and NASA.
This research made use of APLpy, an open-source plotting package for Python \citep{robitaille12},  the colormaps in the CMasher package \citep{vandervelden20}, and the legacystamps package (\url{https://github.com/tikk3r/legacystamps}).

\section*{Data Availability}

The observational data underlying this article are available in the \lofar\ Long Term Archive (LTA; \url{https://lta.lofar.eu}) under projects LT10\_010 and LT14\_004 and in the \xmm\ Science Archive (\url{http://nxsa.esac.esa.int}) under ObsIDs 0300210601, 0402190301, 0505210301. Simulation data will be made available upon reasonable request to the authors.



\bibliographystyle{mnras}
\bibliography{library.bib}






\bsp	
\label{lastpage}
\end{document}

%% file: latex_mycommands.txt
\def \eg {e.g.}

\def \ie {i.e.}
\def \cf {cf.}

\def \omegam {{\hbox{$\Omega_{\rm m}$}}}
\def \omegal {{\hbox{$\Omega_\Lambda$}}}
\def \hzero {{\hbox{$H_0$}}}
\def \arcmin {\hbox{$^\prime$}}
\def \arcsec {\hbox{$^{\prime\prime}$}}
\def \deg {\hbox{$^\circ$}}

\def \compr {{\hbox{$\mathcal{C}$}}}

\def \msun {\hbox{${\rm M_\odot}$}}

\def \mfive {\hbox{$M_{500}$}}

\newcommand{\ergscmsq }{\mbox{erg s$^{-1}$ cm$^{-2}$}}

\newcommand{\kmsmpc }{\mbox{km s$^{-1}$ Mpc$^{-1}$}}

\newcommand{\mujyb }{\mbox{$\mu$Jy beam$^{-1}$}}

\newcommand{\muG }{\mbox{$\mu$G}}
\newcommand{\whz }{\mbox{W Hz$^{-1}$}}

\newcommand{\obsid }{ObsID}

\newcommand{\uv }{\textit{uv}}

\newcommand{\wsclean }{\textsc{WSClean}}

\newcommand{\esas }{\textsc{esas}}
\newcommand{\esasE }{Extended Source Analysis Software}
\newcommand{\sas }{\textsc{sas}}
\newcommand{\sasE }{Scientific Analysis System}

\newcommand{\pyproffit }{\textsc{pyproffit}}

\newcommand{\xmm }{{\em XMM-Newton}}
\newcommand{\chandra }{{\em Chandra}}

\newcommand{\planck }{{\em Planck}}

\newcommand{\lofar }{LOFAR}
\newcommand{\lofarE }{LOw Frequency ARray}

\newcommand{\meerkat }{MeerKAT}

\newcommand{\lotss }{LoTSS}
\newcommand{\lotssE }{LOFAR Two-meter Sky Survey}

\newcommand{\nvss }{NVSS}
\newcommand{\nvssE }{NRAO VLA Sky Survey}

\newcommand{\desi }{DESI Legacy Imaging Surveys}